\documentclass{elsart}
\usepackage{graphics}
\begin{document}
\begin{frontmatter}
\textbf{Preprint MPI~H-V1-2001}\\
\title{On the size of hadrons}
\author{Hans-Christian Pauli}
\address{Max-Planck Institut f\"ur Kernphysik, D-69029 Heidelberg, Germany}
\and
\author{Asmita Mukherjee}
\address{Saha Institute of Nuclear Physics, Calcutta, India}
\date{29 th May 2001}
\begin{abstract}
    The form factor and the mean-square radius of the pion are 
    calculated analytically from a parametrized form of a 
    $q\bar q$ wave function.
    The numerical wave function was obtained previously 
    by solving numerically an eigenvalue equation for the pion
    in a particular model.
    The analytical formulas are of more general interest 
    than just be valid for the pion 
    and can be generalized to the case with unequal quark masses.  
    Two different parametrizations are investigated. 
    Because of the highly relativistic problem,
    noticable deviations from a non-relativistic formula
    are obtained.
\end{abstract}
\maketitle
\end{frontmatter}
%
%
\section{Introduction}
Hadrons are composite particles and therefore have a size. 
A quantitative measure of this size is the mean-square radius
whose experimental value for the pion ($\pi^+$) is \cite{am84}
$\sqrt{<r^2>}= 0.67\pm 0.02$~fm. 
One determines it by first measuring the electro-magnetic 
form factor $F(Q^2)$ for sufficiently small values of the 
(Feynman-four-) momentum transfer $Q^2=-(p_{e}-p_{e'})^2$, 
and then taking the derivative at sufficiently small $Q^2$, 
{\it i.e.}
\begin{equation}
   <r^2> = \left.-6\ \frac{dF(Q^2)}{dQ^2}\right\vert_{Q^2=0} 
.\label{eq:1}\end{equation}
The form factor can also be calculated. 
One of the most remarkable simplicities of the light-cone formalism is 
that one can write down exact expressions for the 
electro-magnetic form factors.  
As was first shown by Drell and Yan \cite{dry70},  
it is advantageous to  choose a special coordinate frame 
to compute form factors and other current matrix elements
at space-like photon momentum. 
In the Drell frame \cite{leb80}, 
the photon's momentum is transverse to the momentum of the incident hadron
and the incident hadron can be directed along the $z$ direction.
With such a choice the four-momentum transfer is
$ -q_\mu q^\mu \equiv Q^2 = \vec q _{\!\perp} ^{\;2}$,
and the quark current can neither create pairs nor  
annihilate the vacuum \cite{BroPauPin98}.   
The space-like form factor for a hadron 
is just a sum of overlap integrals analogous to 
the corresponding non-relativistic formula \cite{dry70}: 
\begin{eqnarray} 
   F _{S\rightarrow S^\prime}(Q^2) &=&
   \sum_{n} \sum_f e_f \int d[\mu_n]
   \,\Psi_{n,S^\prime}^\star (x_i,\vec \ell_\perp{}_i,\lambda_i)
   \,\Psi_{n,S} (x_i,\vec k_\perp{}_i,\lambda_i) 
,\label{eq:2}\\ \mathrm{with}\quad
   \vec \ell _{\!\perp i} &\equiv& 
   \cases{ \vec k _{\!\perp i} - x_i \vec q _{\!\perp}  
         + \vec q _{\!\perp},   & \hbox{for the struck quark,} \cr
           \vec k _{\!\perp i} - x_i \vec q _{\!\perp},      
         & \hbox{for all other partons.}\cr}
\nonumber\end{eqnarray} 
The $d[\mu_n]$ symbolize the convolution over all momentum 
and helicity space arguments of every $\Psi_{n} $ and
$e_f$ is the charge of the struck quark.
This holds for any (composite) hadron and any initial or final spins $S$,
but is particularly simple for a spin-zero hadron like a pion. 
The wave functions $\Psi_{n}=\Psi_{q\bar q},\Psi_{q\bar q g},\dots$ 
are the Fock-space projections of the eigenstate 
which for a meson for example are 
$ \vert\Psi_{meson}\rangle = \sum_{i} ( 
   \Psi_{q\bar q}(x_i,\vec k_{\!\perp_i},\lambda_i)  \vert q\bar q\rangle   +
   \Psi_{q\bar q g}(x_i,\vec k_{\!\perp_i},\lambda_i)\vert q\bar q g\rangle +
   \dots)$. 
Their computation is the aim of the light-cone approach 
to the bound-state problem in gauge theory \cite{BroPauPin98}, 
by solving $H_{LC}\vert\Psi\rangle = M^2\vert\Psi\rangle$,
with the eigenvalues $M^2$ being the invariant mass-squares 
of the physical mesons.

Strictly speaking, the above expression for the form factor contains
contributions from all Fock space sectors. In this work, we restrict
ourselves to the lowest Fock space projection consisting of a $q {\bar q}$
pair. There exists a nonzero probability of finding the pion in its valence
state, which can be calculated \cite{mac}. 
It is known empirically, that the form factor at low $Q^2$ has 
essentially monopole structure \cite{am84}. 
The mean-square radius is essentially all the information there is
for low $Q^2$. 
In fact, for the nucleon also, there are various effective 
three quark light-cone descriptions. 
In these models, without an explicit form of the effective
potential in the light-cone Hamiltonian for three-quarks, one proceeds with
an ansatz for the momentum space wave function. Both exponential
\cite{BroPauPin98}
and power law \cite{pow,BroPauPin98} falloff of this wave function at large 
$k_\perp$ have been used. The low
$Q^2$ properties of the nucleon like the proton magnetic moment $\mu_p$
and its axial coupling $g_A$ have been investigated in these models 
\cite{BroPauPin98}. It is reasonable to assume that the contributions from
the higher Fock components will only refine this initial approximation
\cite{ref}. 

We take the starting expression as,
\begin{equation} 
   F (Q^2) =
   \!\!\int\!\!dx\,d^2\vec k_{\!\perp}\!\!\left(
   e_1\psi(x,\vec k_{\!\perp}+(1-x)\vec q _{\!\perp})+
   e_2\psi(x,\vec k_{\!\perp}-  x  \vec q _{\!\perp}) \right)
   \psi(x,\vec k_{\!\perp}) 
,\label{eq:4}\end{equation}
for the purpose of calculating its theoretical mean-square radius. 

The relation involves only the $L_z=S_z=0$ component of the
general $u\bar d$ wave function, where
$\psi(x,\vec k_{\!\perp})\equiv
 \Psi_{u\bar d}(x,\vec k_{\!\perp};\uparrow\downarrow)$
is the (normalized) probability amplitude for finding 
the quarks with anti-parallel helicities, particularly
for finding the $u$-quark with 
longitudinal momentum fraction $x$ and 
transversal  momentum $\vec k_{\!\perp}$,
and the $\bar d$ with $1-x$ and $-\vec k_{\!\perp}$.
Their respective charges are $e_1$ and $e_2$, respectively, 
with $e_1+e_2=1$.

The $u\bar d$-component of the pion ($\pi^+$) is available
in numerical form, since it has been computed recently 
in the $\uparrow\downarrow$-model \cite{Pau99b}.
But the three-dimensional numerical integration of Eq.(\ref{eq:4}) 
and it subsequent derivation with respect to $Q^2$ 
is cumbersome and may be numerically inaccurate.
The aim of the present work is therefore to calculate  
the mean-square radius $<r^2>$ 
analytically by a suitable parametrization of the 
numerical wave function $\psi(x,\vec k_{\!\perp})$.
The general procedure outlined in the subsequent sections
is applicable also to more general cases.

\section{General considerations}
Usually, one is able to write down an integral
equation in the three variables $x$ and $\vec k _{\!\perp}$ for 
the wavefunction $\psi(x,\vec k _{\!\perp})$
\cite{Pau99b}.
The solution of such an equation is numerically nontrivial,
among other reasons, because the longitudinal momentum fractions 
are limited to $0\leq x \leq 1$.
It is therefore advantageous to substitute the integration variable $x$ 
by another variable $-\infty\leq k_z\leq \infty$ 
which has the same range than either of the two transversal momenta
$\vec k _{\!\perp}$.  For unequal quark masses, the transformation is
defined as,
\begin{equation}
x(k_z)={E_1+k_z\over {E_1+E_2}}
\end{equation}
with $E_{1,2}=\sqrt {{\bar m}_{1,2}+k_\perp^2+k_z^2}$. In the lowest order
approximation, the effective quark masses ${\bar m}_{1,2}=m_{1,2}$.

\begin{figure}[t]
\begin{minipage}[t]{66mm}
  \resizebox{1.00\textwidth}{!}{\includegraphics{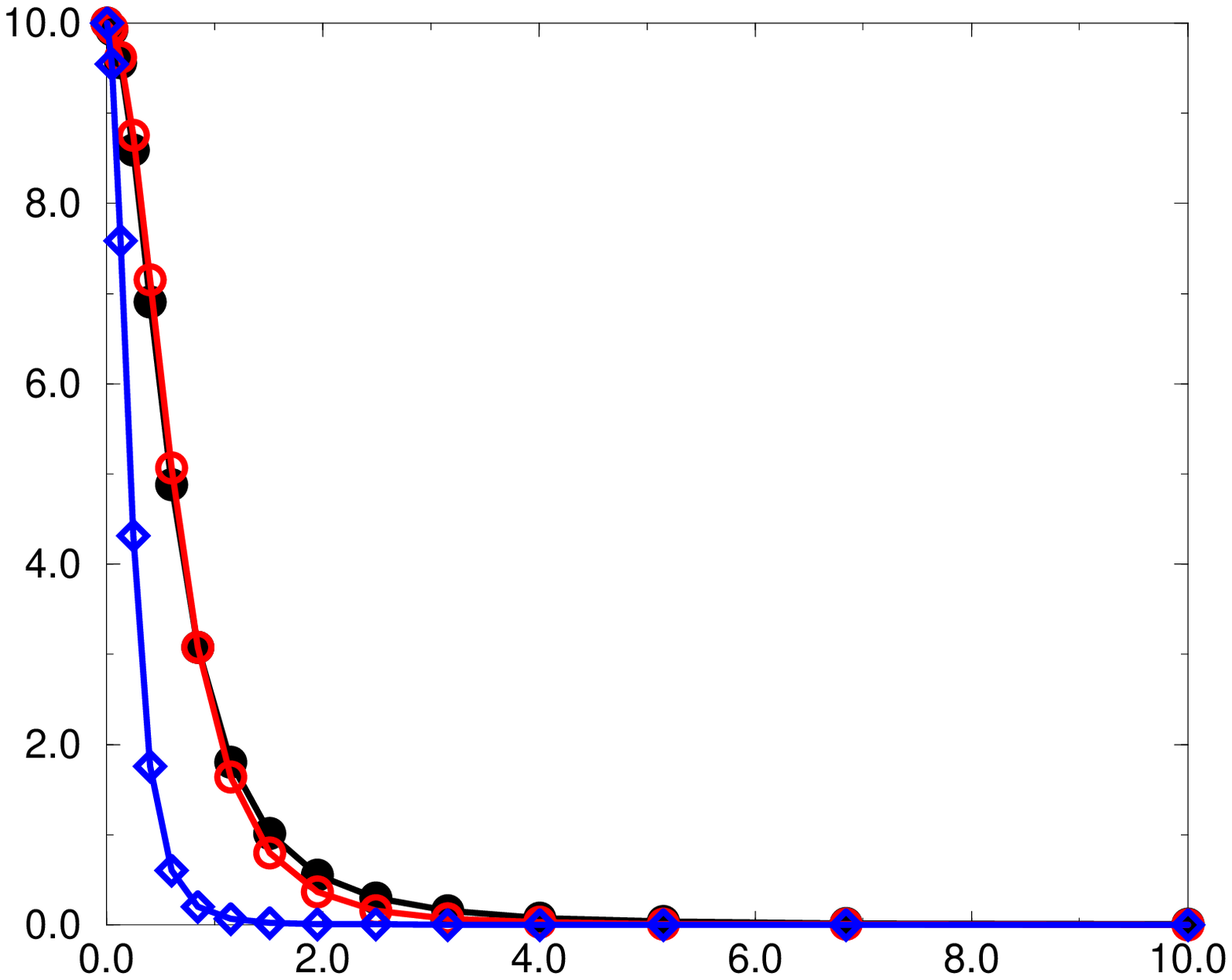}}
  \caption{\label{fig:b1}   
   The pion wave function $\Phi(p)$ 
   is plotted versus $p/(1.552\kappa)$
   in an arbitrary normalization. 
   The filled circles indicate the numerical results,
   the open circles the fit function.
   The diamonds denote the pure Coulomb solution. 
}\end{minipage}
\ \hfill
\begin{minipage}[t]{66mm}
  \resizebox{0.95\textwidth}{!}{\includegraphics{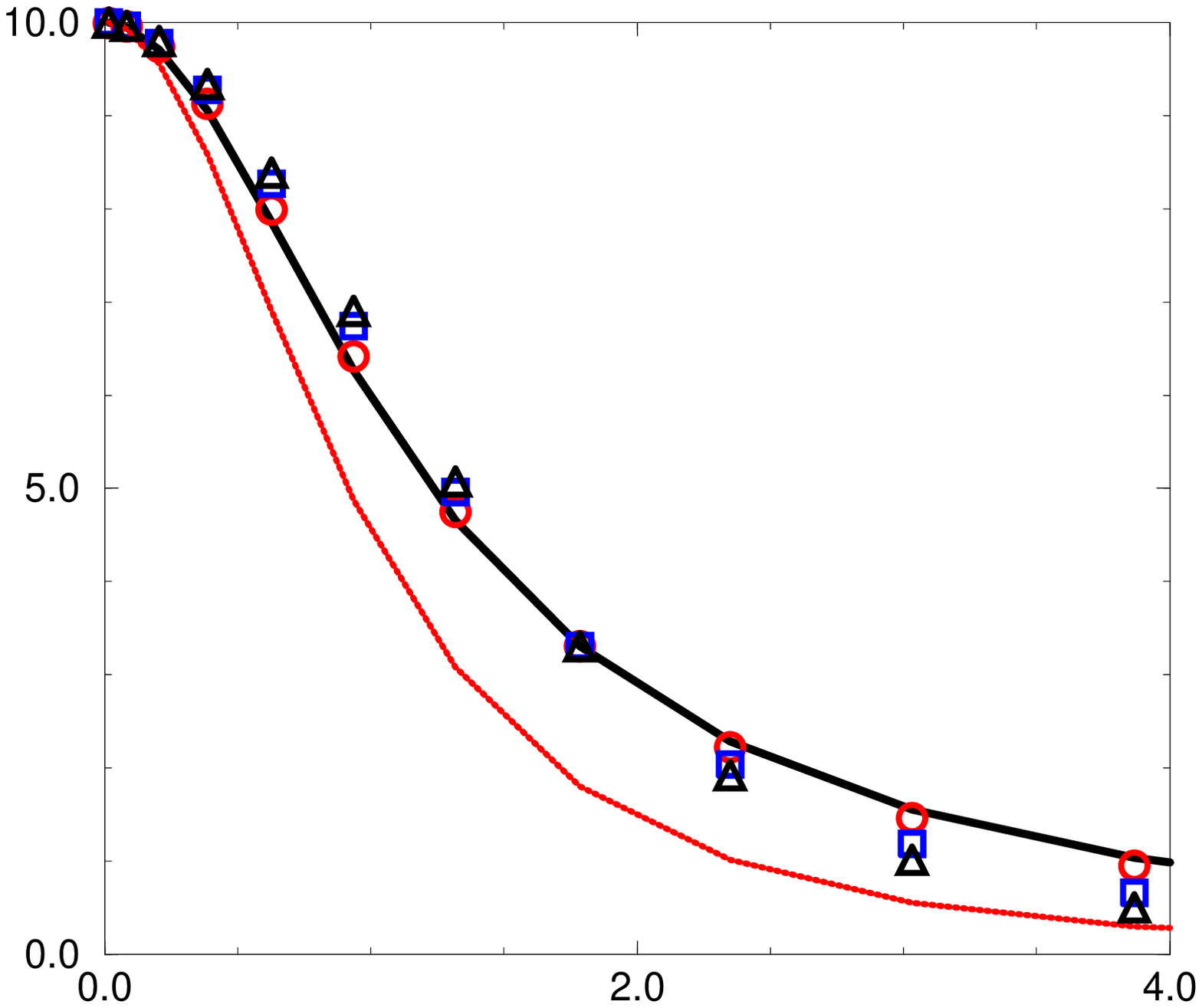}}
  \caption{\label{fig:b2} 
   The function $\chi(p)$ is plotted versus p and 
   fitted to the three exponents $\kappa=2,\frac{3}{2},1$
   (triangle,box,circle).  
   To guide the eye, $\varphi$ is included by the dotted line.
   Note the increased scale as compared to the left.
}\end{minipage} 
\end{figure}

For equal quark masses $m_u=m_d=m$, 
the substitution is so simple that it can even be inverted, {\it i.e.}
\begin{equation}
   x(k_z) = \frac{1}{2}\left(1 +
   \frac{k_z}{\sqrt{m^2 + \vec k_{\!\perp}^{\,2} + k_z^2}}\right)
\quad\Longleftrightarrow\quad
   k_z^2 = (m^2+\vec k_{\!\perp}^{\,2}) 
   \frac{(x-\frac{1}{2})^2}{x(1-x)}
.\label{eq:xkz}\end{equation}
Formally, the three integration variables $k_z$ and $k_{\!\perp}$
look like a conventional 3-vector $\vec p \equiv (k_z,\vec k_{\!\perp})$. 
If one substitutes, in addition, the unknown function
$\psi(x,\vec k _{\!\perp})$ by an other unknown function 
$\varphi(\vec p)$ according to
\begin{equation}
   \psi(x,\vec k _{\!\perp}) = \varphi(k_z,\vec k _{\!\perp})
   \ \frac{N}{\sqrt{x(1-x)}} 
   {\left(1+\frac{\vec p ^{\;2}}{m^2}\right)^{\frac{1}{2}}}
,\label{eq:psi}\end{equation}
one gets an identical integral equation in the three variables $\vec p$,
which looks like an integral equation in usual momentum space.
In the $\uparrow\downarrow$-model \cite{Pau99b},
the integral equation is further simplified 
and at the end looks very simple indeed,
\begin{eqnarray*}
   M^2\varphi(\vec p) &=& \left[4m^2+4\vec p^{\,2}\right]
   \varphi(\vec p)
\\ &-& 
   \frac{4}{3}\alpha\ \frac{1}{2\pi^2}\int\frac{d^3\vec p\,'}{m} 
   \left(\frac{4m^2}{(\vec p - \vec p\,')^{\,2}} + 
   \frac{2\mu^2}{\mu^2+(\vec p - \vec p\,')^{2}} \right)
   \varphi(\vec p\,')
.\nonumber\end{eqnarray*}
The equation was solved numerically in \cite{Pau99b} 
for spherical symmetry $\varphi(\vec p)=\varphi(p)$ 
and for the parameter values
$m=1.16$, $\mu=3.8$, and $\alpha = 0.6904$, 
where masses (and momenta) are expressed in units of
$T = 350\ \mathrm{MeV}$. 
The calculated eigenvalue for the ground state agrees with the 
pion mass-squared to a high degree of accuracy, and is stable
with respect to changes in $\mu$ (renormalization).
As can be seen from Fig.~\ref{fig:b1}, it behaves like a power law
$\varphi(p) \sim (1+(p/p_a)^2)^{-\kappa}$,
rather than as anticipated in \cite{leb80}
like a Gaussian $\varphi(p) \sim \exp(-(p/p_g)^2) $.
As shown in Appendix.~\ref{app:a}, a value of $\kappa=2$ is more likely
than others and a fit of $p_a$, 
\begin{equation}
   \varphi(p) = \left(\frac{1}{1+p^2/p_a^2}\right)^{2}
   ,\qquad p_a=1.471
,\label{eq:ff}\end{equation}
reproduces the numerical wave
function quite well as shown in Fig.~\ref{fig:b1},
with albeit a comparatively large Bohr momentum $p_a$.
Expressing the latter in a length, 
the value of $\sqrt{3}/p_a = 0.66$~fm 
(similar to the experimental rms) is no numerical co-incidence, 
but the result of a constraint on the procedure in \cite{Pau99b}. 
 
One should emphasize that only $\psi(x,\vec k _{\!\perp})$ 
but not $\varphi(p)=\varphi(x,\vec k _{\!\perp})$ 
is a probability amplitude, and that the two can
differ appreciably from each other according to Eq.(\ref{eq:psi}),
particularly for a large Bohr momentum $p_a\sim m$ as in 
Eq.(\ref{eq:ff}). Disregarding this proviso,
and mistaking $\varphi(p)$ as a probability
amplitude for finding the particle with relative momentum $p$,
the calculation of the rms-radius would be trivial:
One takes the Fourier transform of Eq.(\ref{eq:ff}),
{\it i.e.} $\mbox{FT}[\varphi(p)]\simeq\exp{(-p_a r)}$,
and calculates its second moment to be 
\[
   <r^2>_{FT} = \frac
   {\int dr\,r^4\,\exp{(-2p_ar)}}
   {\int dr\,r^2\,\exp{(-2p_ar)}} = 
   \frac{3}{p_a^2}
.\]
However, as the variable conjugate to the relative momentum $p$,
the quantity $r$ is the \textit{relative distance}
of the particles. It is \textit{not the radius-distance} from the common
center-of-mass, with respect to which the rms is usually calculated. 
The latter is $r/2$ for equal mass particles, thus
\begin{equation}
   <r^2>_{nr}  =  \frac{<r^2>_{FT}}{4} = \frac{3}{4p_a^2}
.\label{eq:7}\end{equation}
It will be referred to as the `non-relativistic estimate',
since the above construction holds approximately
if $k_z^2\ll m^2$ and $k _{\!\perp}^{\;2}\ll m^2$,
thus $x\sim \frac{1}{2}$ according to Eq.(\ref{eq:xkz}).
The question we pursue in the present work is thus:
How large is the discrepancy between the 
non-relativistic estimate of Eq.(\ref{eq:7})
and the quasi-exact form factor according to Eq.(\ref{eq:4})
and its behaviour for low $Q^2$ according to Eq.(\ref{eq:1}),
particularly for large Bohr momenta.

Before proceeding with the computation of the
form factor, a number of notational definitions 
will be introduced, in terms of which the final
results turn out to be simple.
Once one has $\varphi(\vec p)$ 
in a parametrized form like Eq.(\ref{eq:ff})
one can transform back to the variables $x$ and $\vec k _{\!\perp}$.
Since $\vec p ^{\;2} \equiv k_z^{2} + \vec k _{\!\perp}^{\;2}$
one can use Eq.(\ref{eq:xkz}) to get 
\begin{equation}
   1 + \frac{p^2}{p_a^2}= 1+s^2
   \frac{(x-\frac{1}{2})^2} {x(1-x)} +
   \frac{s^2}{4 m^2 }
   \frac{\vec k_{\!\perp}^{\,2}} {x(1-x)} \equiv
   Z(x,\vec k_{\!\perp}^{\,2})
.\label{eq:Z}\end{equation}
The dimensionless variable $s$ is introduced conveniently as
\begin{equation}
   s=\frac{m}{p_a}
,\label{eq:s}\end{equation}
as well as the isolation of the pure $x$-dependence by
\begin{equation}
   X(x) \equiv 1+s^2 \frac{(x-\frac{1}{2})^2} {x(1-x)}
.\label{eq:xox}\end{equation}
The combination $1+p^2/m^2$ is then trivially obtained by $s=1$.
One can thus compute the form factor according to Eq.(\ref{eq:4}),
{\it i.e.}
\begin{eqnarray}
   F(q_{\!\perp}^2) &=& \int dx\,dk_{\!\perp}^2\,
   \vert\psi(x,\vec k_{\!\perp})\vert^2 \,
   f(x,k_{\!\perp};q_{\!\perp})
,\\
   <r^2> &=& \int dx\,dk_{\!\perp}^2\,
   \vert\psi(x,\vec k_{\!\perp})\vert^2 \,g(x,k_{\!\perp})
.\label{eq:15}\end{eqnarray}
The function $f(x,k_{\!\perp};q_{\!\perp})$ and 
$g(x,k_{\!\perp})$ contain all the difficulty in the problem, 
\begin{eqnarray}
   f(x,k_{\!\perp};q_{\!\perp}) &=&
   \int\limits_0^{2\pi} d\phi\ %
   \frac{e_1\psi(x,\vec k_{\!\perp}+(1-x)\vec q _{\!\perp})
       + e_2\psi(x,\vec k_{\!\perp}-  x  \vec q _{\!\perp})}
   {2\,\psi(x,\vec k_{\!\perp})}
,\label{eq:f}\\
   g(x,k_{\!\perp}) &=& -6 \left.\displaystyle
   \dot f(x,k_{\!\perp};q_{\!\perp})
   \right|_{q_{\!\perp}=0} =-6 \left.\displaystyle
   \frac{d}{d q_{\!\perp}^2} f(x,k_{\!\perp};q_{\!\perp})
   \right|_{q_{\!\perp}=0} 
,\label{eq:g}\end{eqnarray}
that is the integration over the angle $\phi$ 
between $\vec k_{\!\perp}$ and $\vec q_{\!\perp}$
($\vec k_{\!\perp}\vec q_{\!\perp} = k_{\!\perp} q_{\!\perp}\cos\phi$).
Finally the size parameter is introduced,
\begin{equation}
   S \equiv 
   \frac{\sqrt{<r^2>_{cal}}}{\sqrt{<r^2>_{exp}}} =
   \frac{\sqrt{<r^2>_{cal}}}{0.67~\textrm{fm}} 
,\label{eq:size}\end{equation}
as a dimensionless measure of the size. 
\section{The parametrization of the wave function}

Since the quarks move highly relativistically ($p_a^2 \gg m^2$)
and one cannot disregard the factor $\sqrt{1+p^2/m^2}$
in Eq.(\ref{eq:psi}). 
One way to account for that is to parametrize directly
\begin{equation}
   \chi(p) \equiv \varphi(p)
   \sqrt{1+\frac{\vec p ^{\;2}}{m^2}} = 
   \left(1+\frac{\vec p ^{\;2}}{p_s^{\;2}}\right)^{-\kappa} 
,\label{eq:chi}\end{equation}
with the two adjustable parameters $\kappa$ and $p_s$.
We have performed two fits,  for two fixed values of $\kappa$,
and have calculated analytically 
the size parameter $S$ according to Eq.(\ref{eq:size}). 
The results are: 
\[
\begin{tabular}{cccc} 
  $\kappa=2,$ &           
   $s=0.5588,$ & $p_s = 2.075,$ & $S=0.4577,$ \\
   $\kappa=\frac{3}{2},$ & 
   $s=0.6796,$ & $p_s = 1.707,$ & $S=0.4321.$  
\end{tabular}
\]
One needs to consider essentially the integral
\[
   1 = \int dx d^2 \vec k_{\!\perp} |\psi(x,\vec k _{\!\perp})|^2
   \quad
.\]
Inserting $\psi$ according to 
Eqs.(\ref{eq:psi}) and (\ref{eq:chi}) 
gives with $s=m/p_s$
\begin{eqnarray*}
   \frac{1}{N^2} &=& 
   \int\limits_0^1 \frac{dx}{x(1-x)} 
   \int\limits_0^\infty dk_{\!\perp}^{\,2}
   \int\limits_0^{2\pi} \frac{d\phi}{2} 
   \ \frac{1}{\left[Z(x,\vec k_{\!\perp}^{\,2})\right]^{2\kappa}}.
\end{eqnarray*}
The integration over $\phi$ is trivial, 
and the integration over $k_{\!\perp}^{\,2}$ elementary, 
since
\begin{equation}
   \int\limits_0^\infty \frac{dk_{\!\perp}^{\,2}}
   {\left[Z(x,\vec k_{\!\perp}^{\,2})\right]^{\,n}} =
   \frac{4p_s^2x(1-x)}{(n-1)\left[X(x)\right]^{n-1}}
,\label{eq:elem}\end{equation}
with $X(x)$ defined in Eq.(\ref{eq:xox}).
Introducing
\[
   A (s) = \int\limits_0^1 dx
   \frac {1} {\left[X(x)\right]^{2\kappa-1}}
,\]
by definition, one remains with
\begin{equation}
   \frac{1}{N ^2} = \frac{4\pi p_s^2}{2\kappa-1} A (s)
.\label{eq:11}\end{equation}
Here the matter rests, since $A$ cannot be integrated
in closed form. 

For to compute the means-square radius, we first evaluate the
function $f$ as defined in Eq.(\ref{eq:f}).
In the general case one gets
\begin{eqnarray*}
   f(x,k_{\!\perp};q_{\!\perp}) &=& 
   \frac{e_1}{2} Z^\kappa \int\limits_{0}^{2\pi}
   d\phi\ \left[1+\displaystyle s^2\frac{(x-\frac{1}{2})^2}{x(1-x)}+
    \frac{\left(\vec k_{\!\perp}-(1-x)\vec q_{\!\perp}\right)^{\,2}}
    {4 p_s^2 x(1-x)}\right]^{-\kappa}
\nonumber\\ &+&
   \frac{e_2}{2} Z^\kappa \int\limits_{0}^{2\pi}
   d\phi\ \left[1+\displaystyle s^2\frac{(x-\frac{1}{2})^2}{x(1-x)}+
    \frac{\left(\vec k_{\!\perp}-x\vec q_{\!\perp}\right)^{\,2}}
    {4 p_s^2 x(1-x)}\right]^{-\kappa}
.\end{eqnarray*}
Using the abbreviative coefficient functions 
depending on $x$ and $k_{\!\perp}^{\,2}$, {\it i.e.}
\begin{eqnarray}
  &&b_1 = Z+\frac{1-x}{x}\ \frac{q_{\!\perp}^{\,2}}{4p_s^2},\hskip3em
    c_1 = -\frac{1}{x}\frac{k_{\!\perp}}{p_s} \frac{q_{\!\perp}}{2p_s},
\\
  &&\dot b_1 = \frac{1}{4\ p_s^2}    \ \frac{1-x}{x}, \hskip4em
   c_1\dot c_1 = \frac{1}{2\ p_s^2}\ \frac{1-x}{x}\ (Z-X)
,\label{eq:21}\end{eqnarray}
and correspondingly $b_2,c_2,\dot b_2$ and $\dot c_2$
by exchanging $x\leftrightarrow 1-x$, gives
\[
   f(x,k_{\!\perp};q_{\!\perp}) = 
   Z^\kappa\int\limits_{0}^{\pi}d\phi\ \left[
   \frac {e_1}{\left( b_1 + c_1\cos\phi\right)^\kappa}+
   \frac {e_2}{\left( b_2 + c_2\cos\phi\right)^\kappa} \right]
.\]
The integration over $\phi$ can not be carried out 
in closed form for general values of $\kappa$.

\subsection{The integrals for $\kappa=2$}
The normalization integral is now
\[
   \frac{1}{N ^2} = \frac{4\pi }{3} p_s^2 A (s)
.\]
The contribution to $f$ from the up-quark becomes to leading order in $c$ 
\[
   \frac{\dot f}{f} = 
   \frac{\dot b}{b} - 3\frac{b\dot b-c\dot c}{b^2-c^2} \simeq 
   -2\frac{\dot b}{b} + 3 \frac{c\dot c}{b^2}
.\]
Inserting this into the definition of $g$ in Eq.(\ref{eq:g}) gives
\[
   g(x,k_{\!\perp}) =
   \frac{3\pi}{x(1-x)p_s^2} \left[ -
   \frac{2}{Z} + \frac{3X}{Z^2} \right] x^2
.\]
Here we have taken both the quark and antiquark contributions (see Appendix
B).

\begin{figure} [t]
\begin{minipage}[t]{67mm}
  \resizebox{0.99\textwidth}{!}{\includegraphics{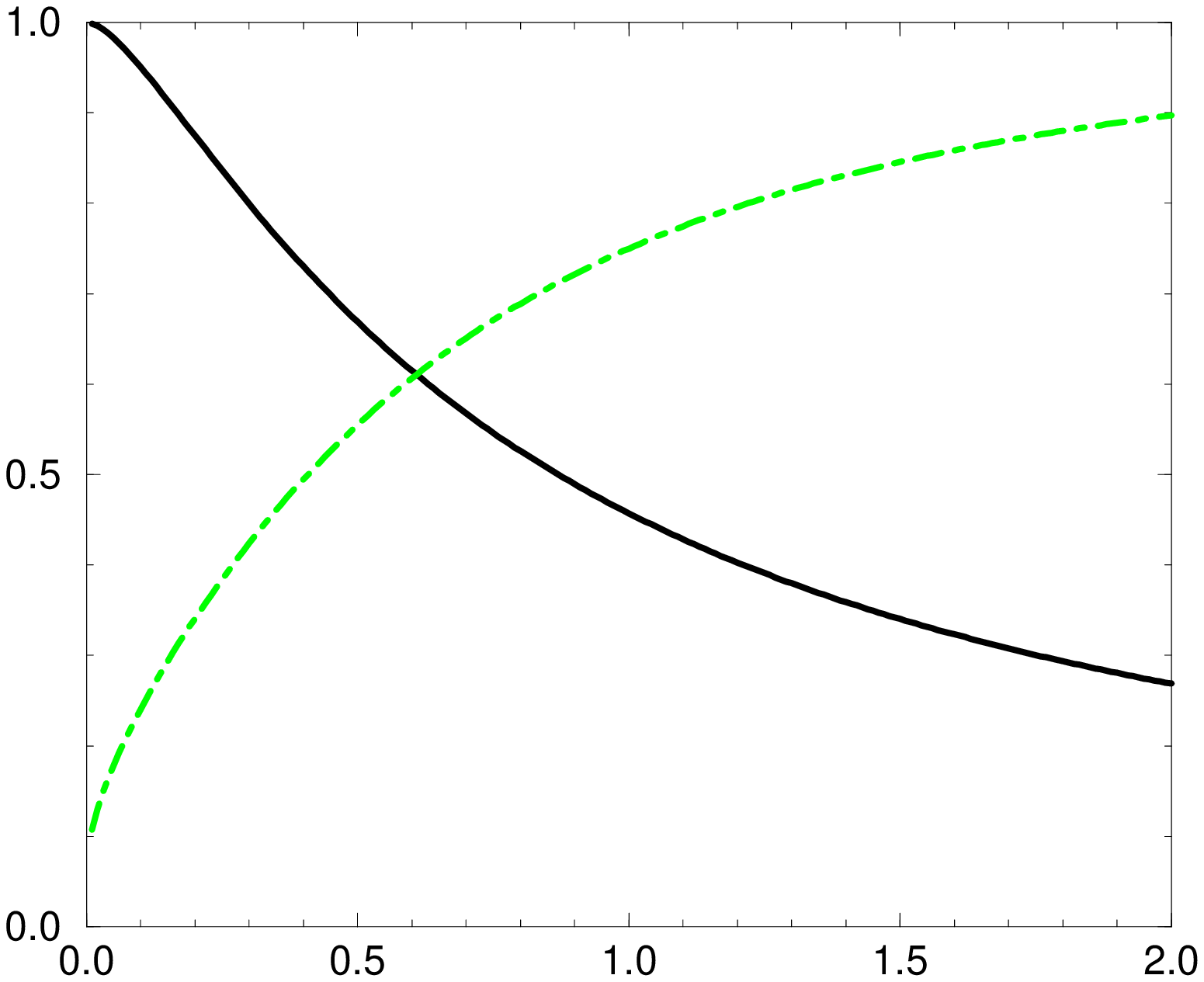}}
  \caption{\label{fig:ACR4} 
    For $\kappa=2$, the functions $A (s)$ and $<r^2>$ (in units of ${3\over
    4p_s^2})$ 
    are plotted versus $s$ by the solid and 
    the dashed-dotted line, respectively.}
\end{minipage}
\ \hfill
\begin{minipage}[t]{67mm}
  \resizebox{0.99\textwidth}{!}{\includegraphics{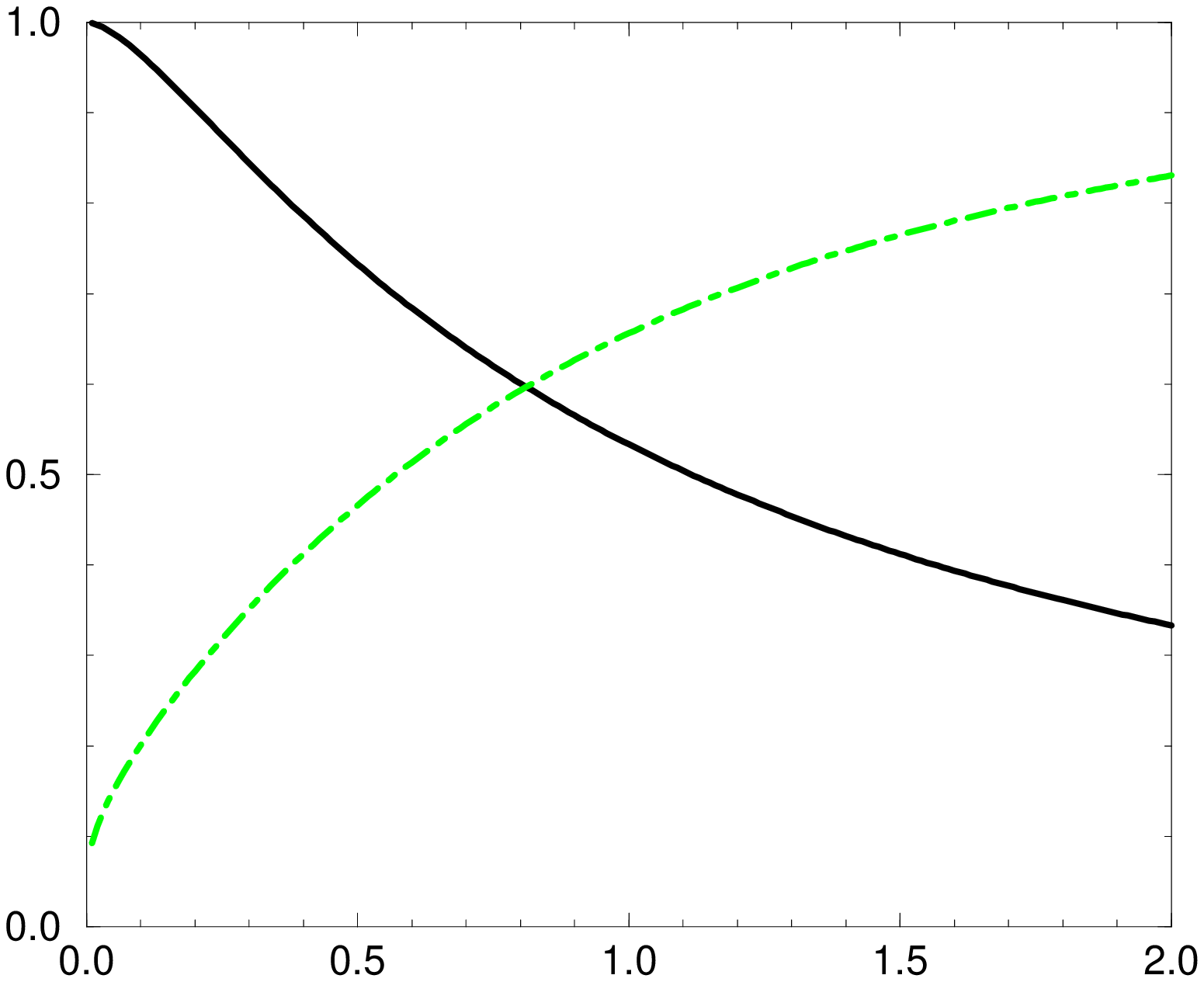}}
  \caption{\label{fig:ACR3} 
    For $\kappa=\frac{3}{2}$, the functions $A (s)$ and 
    $<r^2>$ (in units of ${3\over {4p_s^2}})$ are plotted versus $s$ by the solid 
    and the dashed-dotted line, respectively.}
\end{minipage}
\end{figure}

\begin{figure} [t]
\begin{minipage}[t]{67mm}
  \resizebox{0.99\textwidth}{!}{\includegraphics{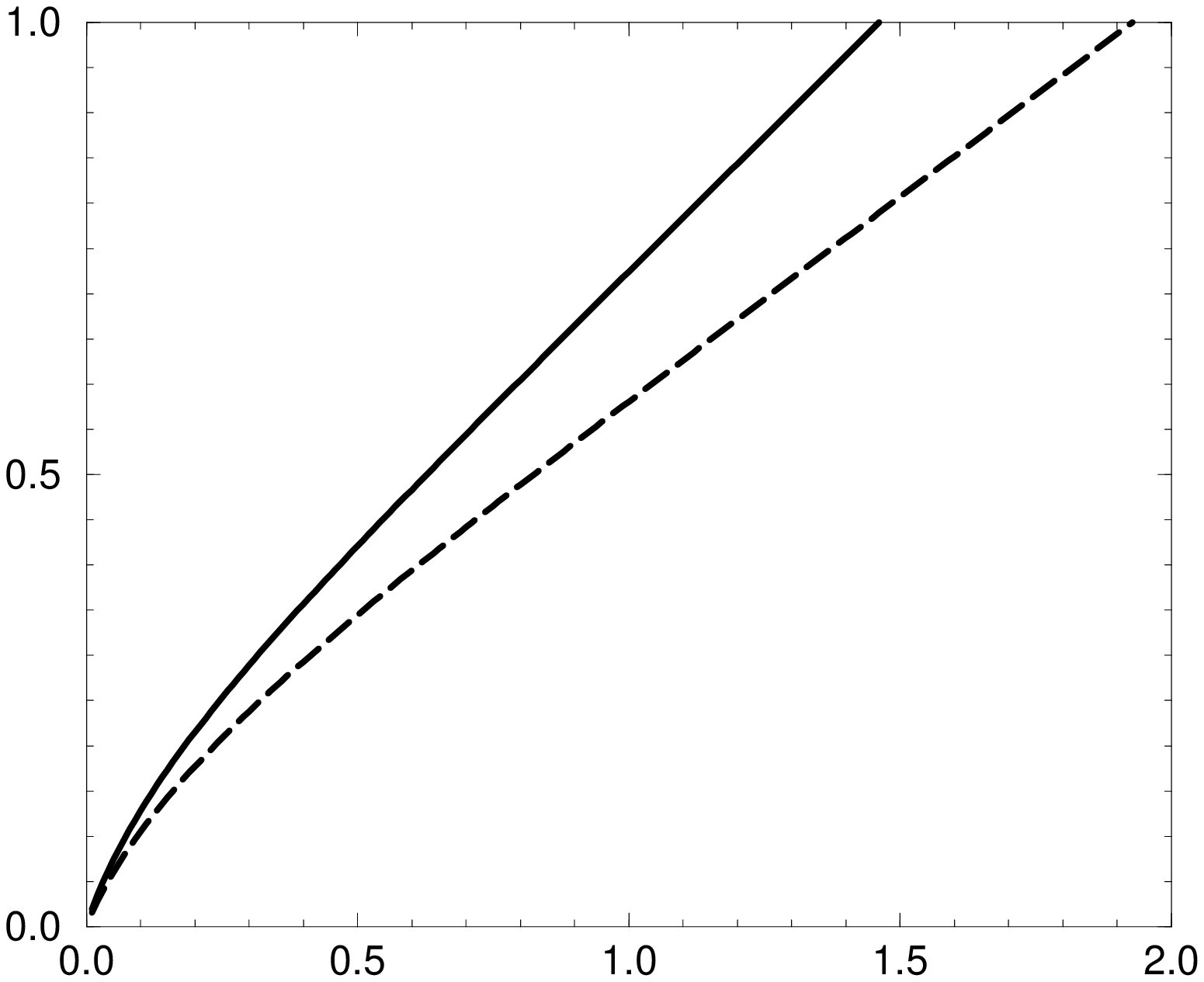}}
  \caption{\label{fig:rms} 
    The root-mean-square radius of the pion,
    ({\it i.e. }$\sqrt{<r^2>}/0.67$~fm), is plotted versus $s$ 
    by the  solid line for $\kappa=2$ and
    by the dashed line for $\kappa=\frac{3}{2}$.}
\end{minipage} 
\ \hfill
\begin{minipage}[t]{67mm}
  \resizebox{0.99\textwidth}{!}{\includegraphics{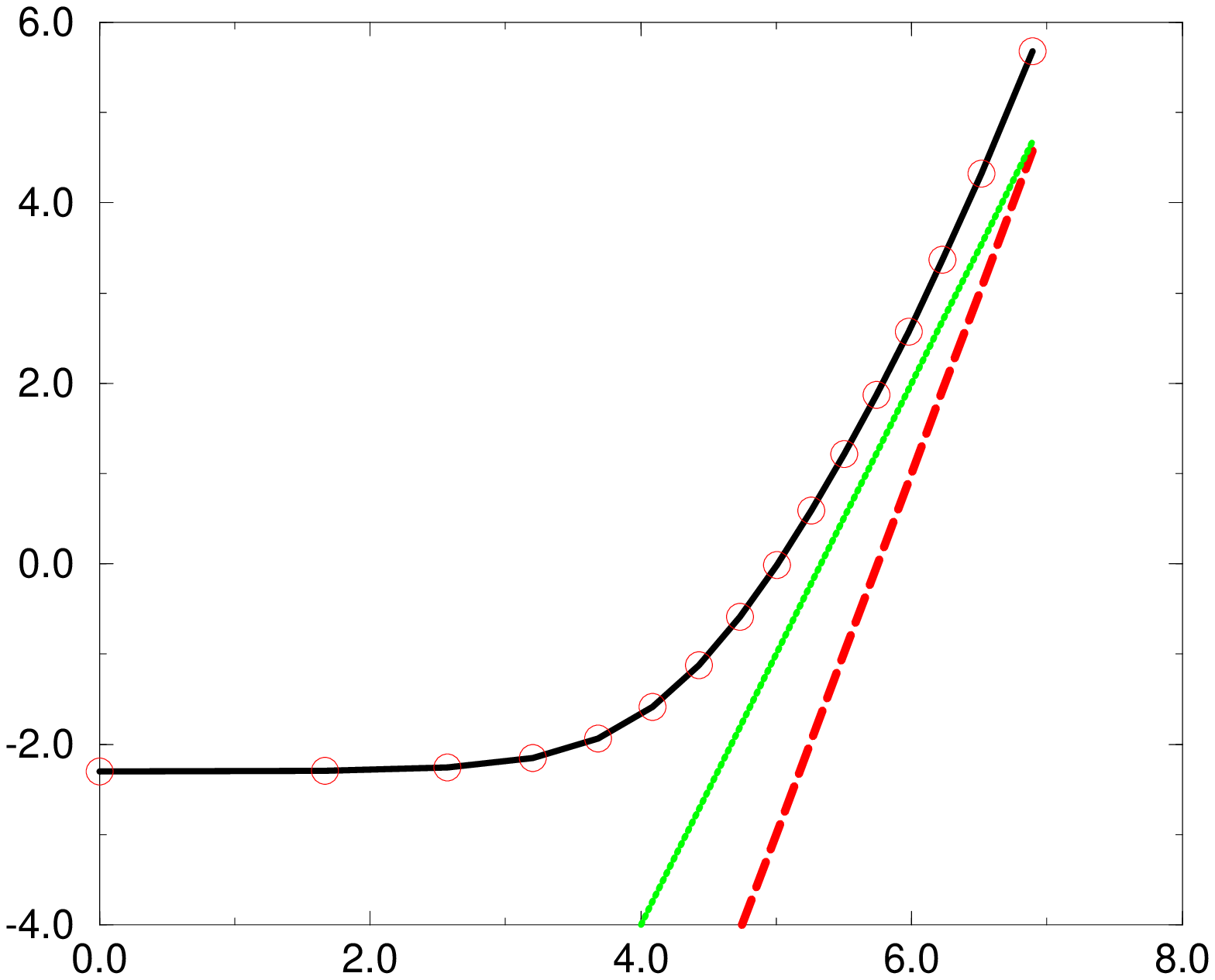}}
  \caption{\label{fig:a1}  
   $\ln{1/\varphi}$ is plotted versus $\ln{p}$ (circle),
   and compared with $4\ln{p}$ (dashed) and 
   $3\ln{p}$ (dotted line). } 
\end{minipage} 
\end{figure}

With the elementary $k_{\!\perp}^2$-integration of 
Eq.(\ref{eq:elem}), one ends up with 
\[
   <r^2> = \frac{6\pi}{5} N ^2\int\limits_0^1 
   \frac{dx}{x(1-x)}
   \frac{x^2}{\left[X(x)\right]^{4}} =
   \frac{9}{10 p_s^2} \frac{R (s)}{A (s)}
.\]
The two elementary integrals $A$ and $R$ are complicated
but straightforward and yield
\begin{eqnarray*}
   A(s) &=& \int\limits_0^1 dx
   \frac{1}{\left[X(x)\right]^{3}} =
   \frac{-8-10s^2+3s^4+3b(s)s^2\left(8 -4s^2 +s^4)\right)}
   {8(s^2-1)^3}
,\\
   R(s) &=& \int\limits_0^1    \frac{dx}{x(1-x)}
   \frac{x^2}{\left[X(x)\right]^{4}} =
   \frac{-136+ 72s^2 -56s^4 + 15 s^6}{48 (s^2-1)^4}
\nonumber\\ && \hskip6em +
   3 b(s)\ \frac{32 -16s^2 +36s^4 -22s^6 +5s^8}{48 (s^2-1)^4}
.\end{eqnarray*}
The auxiliary function $b(s)$ is an abbreviation for 
\begin{equation}
   b(s)= \frac{\mathrm{arctan}(\sqrt{s^2-1})}
     {\phantom{\mathrm{arctan}(}\sqrt{s^2-1}} 
.\label{eq:b(s)}\end{equation}
The functions are plotted in Fig.~\ref{fig:ACR4}.
As seen there, the asymptotic behaviour for $s\longrightarrow\infty$,
\[
   A(s) \rightarrow \frac{3\pi}{16s}, \qquad 
   R(s) \rightarrow \frac{5\pi}{32s} 
,\]
is practically reached for values as small as $s=2$.
Note that in the asymptotic limit $s\rightarrow\infty$,
the mean-square radius reaches the correct 
non-relativistic value  $<r^2>_{nr}=3/(4p^2_s)$. 
As a final result, the size 
\begin{eqnarray*}
   S &=& 
   \frac{\hbar c}{0.67\mbox{\,fm\,}T} 
   \frac{s}{m}\ \sqrt{\displaystyle\frac{9R(s)}{10A(s)}}= 
   \frac{0.8401}{m}\ s\ \sqrt{\displaystyle\frac{9R(s)}{10A(s)}}
,\end{eqnarray*}
is plotted in Fig.~\ref{fig:rms}  
for the mass value $m=1.16$ as function of $s$.

%
\subsection{The case $\kappa=\frac{3}{2}$}
According to Eq.(\ref{eq:11}),
the normalization integral is for $\kappa=\frac{3}{2}$ 
\[
   \frac{1}{N^2} = 2\pi p_s^2 A(s)
.\]
In leading order in $c$, the contribution to the rms from the up-quark 
becomes 
\[
   \frac{\dot f}{f} = 
   - \frac{3}{2}\frac{\dot b}{b} + \frac{15}{8} \frac{c\dot c}{b^2}
.\]
Inserting the derivatives from Eq.(\ref{eq:21}) gives (see Appendix C)
\[
   g(x,k_{\!\perp}) =
   \frac{3\pi}{8p_s^2\,x(1-x)} \left[ -
   \frac{9}{Z} + \frac{15X}{Z^2} \right] x^2
.\]
With the elementary $k_{\!\perp}^2$-integration of 
Eq.(\ref{eq:elem}), one ends up with 
\[
   <r^2> = \frac{9\pi}{8} N^2 \int\limits_0^1 \frac{dx}{x(1-x)}
   \frac{x^2}{\left[X(x)\right]^{3}} =
   \frac{9}{16 p_s^2} \frac{R(s)}{A(s)}
,\]
according to Eqs.(\ref{eq:11}) and (\ref{eq:15}). 
The elementary integrals are now
\begin{eqnarray*}
   A(s) &=& \int\limits_0^1 dx
   \frac{1}{\left[X(x)\right]^{2}} =
   \frac{2+s^2+b(s)s^2\left(-4+s^2\right)}{2(s^2-1)^2}
,\\
   R(s) &=& \int\limits_0^1    \frac{dx}{x(1-x)}
   \frac{x^2}{\left[X(x)\right]^3} =
   \frac{-20-8s^2+3s^4}{8(s^2-1)^3}
\nonumber\\ && \hskip6em +
   \ b(s)\ \frac{-16+8s^2 +10s^4 +3s^6}{8(s^2-1)^3}
.\end{eqnarray*}
The auxiliary function $b(s)$ is the same as in Eq.(\ref{eq:b(s)}).
The functions are plotted in Fig.~\ref{fig:ACR3}.
Finally, the size function
\begin{eqnarray*}
   S &=& \frac{\hbar c}{0.67\mbox{\,fm\,}T} 
   \frac{s}{m}\ \sqrt{\displaystyle\frac{9R(s)}{16A(s)}}= 
   \frac{0.8401}{m}\ s\ \sqrt{\displaystyle\frac{9R(s)}{16A(s)}}
\end{eqnarray*}
is plotted in Fig.~\ref{fig:rms} 
for the mass value $m=1.16$ as function of $s$.
\section{Summary and conclusion}
We have analytically calculated the mean square radius for the pion in the 
light-cone approach in the valence sector. The contributions
from the higher Fock space sectors are expected to only refine this initial
approximation.  We  have used a
parametrized form of the wave function in momentum space which had been
obtained previously by solving the eigenvalue equation for the effective
light-cone Hamiltonian. The mean square radius is calculated by first 
calculating the form factor
at sufficiently low $Q^2$ and then taking its derivative with respect to
$Q^2$ at $Q^2=0$. We have investigated two different
values of the parametrization. $\kappa=2$ is found to be a better fit than
$\kappa={3\over 2}$. We have shown that the mean square radius
deviates noticably from the non-relativistic estimate because of the
relativistic effects. In the asymptotic region (large $m^2$), it approaches the correct
non-relativistic value. 

\section{Acknowledgement}
AM would like to thank A. Harindranath for various useful discussions.

\begin{appendix}
\section{\label{app:a}The numerical wave function of the pion}
The function $\varphi(p)$ was cumputed in previous work \cite{Pau00c}
and is tabulated in Table~\ref{tab:1}. 
It behaves much like an inverse power 
$\varphi(p) \sim (1+(p/p_a)^2)^{-\kappa}$.
The value of $\kappa$ is closer to $2$ than to $\frac{3}{2}$,
as demonstrated in Figure~\ref{fig:a1}.
Unfortunately, the maximum momentum of the Gaussian 
quadratures does not allow for a preciser statement.

\begin{table} [b]
\caption{\label{tab:1}
    The calculated pion wave function $\varphi(p)$. 
}\vskip1em\begin{center}
\begin{tabular}{|cc||cc|} \hline
    \rule[-1em]{0mm}{1em}
        $p$     &   $\phi(p )$    & 
        $p$     &   $\phi(p )$    \\ \hline
    \rule[1em]{0mm}{0.5em}
     0.015739   &  10.00000       &  
     0.083241   &   9.928756      \\ 
     0.205995   &   9.563656      & 
     0.386376   &   8.592389      \\       
     0.628023   &   6.908239      &        
     0.936173   &   4.877784      \\       
     1.318156   &   3.077257      &        
     1.784203   &   1.801300      \\       
     2.348743   &   1.011136      &       
     3.032468   &   0.554441      \\       
     3.867276   &   0.297233      &       
     4.902437   &   0.154397      \\      
     6.223861   &   0.076201      &       
     7.998102   &   0.034485      \\      
    10.62105    &   0.013264      &       
    15.52088    &   0.003404      \\ \hline
\end{tabular}\end{center}
\end{table}

\section{\label{app:b}The function $g(x,k_{\!\perp})$ for $\kappa=2$}
Let us denote the contribution from the up-quark to the form factor 
function simply $f(x,k_{\!\perp};q_{\!\perp})$.
For $\kappa=2$ one gets:
\[
   f(x,k_{\!\perp};q_{\!\perp}) = 
   Z^2\int\limits_{0}^{\pi}
   d\phi\ \frac {e}{\left( b  + c \cos\phi\right)^2} =
   Z^2\pi e \frac {b }{\left( b ^2 - c ^2\right)^{\frac{3}{2}}}
.\]
Its derivative can be calculated quite in general
\[
   \frac{\dot f}{f} = 
   \frac{\dot b}{b} - 3\frac{b\dot b-c\dot c}{b^2-c^2} = 
   \frac{\dot b(-2b^2+c^2)+3bc\dot c}{b(b^2-c^2)}
.\]
In the limit $q_{\!\perp}\rightarrow0$ (thus $c\rightarrow0$)
it becomes to leading order
\[
   \dot f = 
   \left(-2\frac{\dot b}{b} +3 \frac{c\dot c}{b^2}\right) \pi e 
.\]
Inserting the derivatives from Eq.(\ref{eq:21}) 
and taking the contributions from quark and anti-quark gives
one gets thus with  
\[
   g(x,k_{\!\perp}) =
   \frac{3\pi}{p_s^2} \left[ -
   \frac{2}{Z} + \frac{3X}{Z^2} \right]
   \left(e_1\frac{1-x}{x} + e_2\frac{x}{1-x} \right)
.\]
The expression in the round bracket has a contribution 
$(1-2x)$ which is odd under the exchange $x\leftrightarrow 1-x$, 
\[
   e_1\frac{1-x}{x} + e_2\frac{x}{1-x} = 
   \frac{\left[(e_1+e_2)x^2 + e_1(1-2x)\right]}{x(1-x)} \longmapsto
   \frac{(e_1+e_2)x^2}{x(1-x)}
.\]
It will vanish in the final integration over $x$ and can be omitted.
This leaves one with the final expression
\begin{equation}
   g(x,k_{\!\perp}) =
   \frac{3\pi}{x(1-x)\ p_s^2} \left(e_1 + e_2\right)
   \left[ - \frac{2}{Z} + \frac{3X}{Z^2} \right] x^2 
.\label{eq:g4}\end{equation}

\section{\label{app:c}The function $g(x,k_{\!\perp})$ for $\kappa=\frac{3}{2}$}
The form factor function for the up-quark 
\[
   f = \int\limits_0^\pi d\phi \frac{eZ^\frac{3}{2}}
   {(b + c\,\textrm{cos}\phi)^\frac{3}{2}} =
   \frac{2eZ^\frac{3}{2}}{p}\mathrm{E}[q]
   ,\ \textrm{with}
   \ \cases{p=(b-c)\sqrt{(b+c)},   & \cr
          q=\displaystyle\frac{2c}{b+c},   & \cr}
\]
can be integrated in closed form also for $\kappa={3}/{2}$ 
and be expressed in terms of
the complete elliptic integrals of the second kind, 
$E[q]\equiv\int_0^{\frac{\pi}{2}}d\theta\sqrt{1-q\ \mathrm{sin}^2\theta}$. 
Since ${dE[q]}/{dq} = \left(E[q]-K[q]\right)/{2q}$,  
the derivative is
\[
   \frac{\dot f}{f} = \frac{\dot q}{q}
   \ \frac{E[q]-K[q]}{2E[q]} - \frac{\dot p}{p}
.\]
Both derivatives $\dot q$  and $\dot p$ 
have a potentially dangerous singularity $b\dot c$,
\[
   \frac{\dot q}{q} = \frac{-c\dot b+b\dot c}{(b+c)c}
   ,\qquad\mbox{and}\qquad 
   \frac{\dot p}{p} = \frac{(3b+c)\dot b -(3c+b)\dot c }{2(b^2-c^2)}
.\]
To cure the problem, one must expand 
\[
   \frac{E[q]-K[q]}{E[q]}= -\frac{q}{2}-\frac{5q^2}{16}-\mathrm{O}[q^3]
   \simeq -\frac{c}{b+c}\left(1 + \frac{5}{4}\frac{c}{b+c}\right) 
,\]
to a sufficiently high order, which upon insertion yields
\begin{eqnarray*}
   \frac{\dot f}{f} &=& 
   \dot b \left[\frac{+c}{2(b+c)^2}\left(1+\frac{5c}{4(b+c)}\right)
   - \frac{3b+c}{2(b^2-c^2)}\right]
\nonumber\\ &+&
   \dot c \left[\frac{-b}{2(b+c)^2}\left(1+\frac{5c}{4(b+c)}\right)
   + \frac{b+3c}{2(b^2-c^2)}\right]
.\end{eqnarray*}
The leading terms in the coefficient of $\dot c$ now tend to cancel, 
{\it i.e.}
\begin{eqnarray*}
   \frac{\dot f}{f} &=& 
   \dot b \left[\frac{+c}{2(b+c)^2}\left(1+\frac{5c}{4(b+c)}\right)
   - \frac{3b+c}{2(b^2-c^2)}\right]
\nonumber\\ &+&
   \frac{c\dot c}{b+c}\left[\frac{-5b}{8(b+c)^2} + \frac{3}{2(b-c)}
   + \frac{b}{(b^2-c^2)}\right]
   \longrightarrow 
   - \frac{3}{2}\frac{\dot b}{b} + \frac{15}{8} \frac{c\dot c}{b^2}
.\end{eqnarray*}
In the last step only terms were kept which survive in 
the limit $c\rightarrow0$.
Inserting the derivatives from Eq.(\ref{eq:21}) gives 
for the complete amplitude
\begin{equation}
   g(x,k_{\!\perp}) =
   \frac{3\pi}{8x(1-x)p_s^2} \left(e_1+ e_2\right)
   \left[ - \frac{9}{Z} + \frac{15X}{Z^2} \right] x^2
,\label{eq:g3}\end{equation}
where the same simplifications have been done as in the previous section.
\end{appendix}

\end{document}